\documentclass[preprint2,trackchanges]{aastex6}

\newcommand\pname{K2-98\,b}
\newcommand\ssname{K2-98}
\newcommand\sname{EPIC\,211391664}

\newcommand\corot{\emph{\it CoRoT}}
\newcommand\kepler{\emph{{\it Kepler}}}
\newcommand\vsini{$v$\,sin\,$i_\star$}   
\newcommand\visi{$v$\,sin\,$i_\star$}
 
\newcommand\vmic{$V_{\rm mic}$}
\newcommand\vmac{$V_{\rm mac}$}

\newcommand\teff{$T_{\rm eff}$}

\newcommand\logg{log\,{\it g$_\star$}}
\newcommand\met{[M/H]}
\newcommand\Msun{\hbox{$M_{\odot}$}}  
\newcommand\Rsun{\hbox{$R_{\odot}$}}  
\newcommand\kms{\hbox{km\,s$^{-1}$}}  

\newcommand{\smass}[1][$M_{\odot}$]{$ 1.074\pm0.042$ #1}
\newcommand{\sradius}[1][$R_{\odot}$]{$ 1.311 ^{+ 0.083} _{ - 0.048} $ #1}
\newcommand{\tzero}[1][days]{ $7145.9807 \pm 0.0012$ #1}
\newcommand{\porb}[1][days]{$10.13675 \pm { 0.00033}$ #1}
\newcommand{\ec}[1][]{ $0$ #1 }
 
\newcommand{\inclination}[1][deg]{$ 89.0^{ + 0.5}_{ - 0.7}$ #1}
\newcommand{\saxis}[1][]{$ 15.388^{+0.543}_{-1.192}$ #1} 
\newcommand{\spradius}[1][]{$ 0.0301^{+0.0004}_{-0.0003}$ #1} 
 \newcommand{\qone}[1][]{$ 0.27^{ + 0.29}_{ - 0.12}$ #1}    
\newcommand{\qtwo}[1][]{$ 0.47^{ + 0.26}_{ - 0.24}$ #1} 
 \newcommand{\qonefixed}[1][]{$ 0.40 \pm 0.05$ #1}    
\newcommand{\qtwofixed}[1][]{$ 0.26 \pm 0.05$ #1} 
\newcommand{\krv}[1][m\, s$^{-2}$]{ $ 9.1 \pm 2.3 $ #1 }
\newcommand{\velF}[1][km\, s$^{-2}$]{$ 76.6116 \pm 0.0029 $ #1 }
\newcommand{\velH}[1][km\, s$^{-2}$]{$ 76.7417 \pm 0.0026 $ #1 }
\newcommand{\velS}[1][km\, s$^{-2}$]{$ 76.7479 \pm 0.0022 $ #1 }
\newcommand{\pradius}[1][$R_{\oplus}$]{$4.3^{+0.3}_{-0.2}$ #1}
\newcommand{\axis}[1][AU]{$ 0.0943^{ + 0.0061}_{ - 0.0052} $ #1} 
\newcommand{\impactp}[1][]{$ 0.27^{ + 0.17}_{ - 0.14} $ #1} 
\newcommand{\ttotal}[1][hours] {$ 5.03^{ + 0.05}_{ - 0.04}$ #1 }
\newcommand{\tineg }[1][hours] {$ 0.16^{ + 0.03}_{ - 0.01}$ #1 }
\newcommand{\sden }[1][g\,cm$^{-3}$]{$ 0.66^{ + 0.07}_{ - 0.12} $ #1 }
\newcommand{\uone}[1][]{ $ 0.47^{ + 0.14}_{ - 0.17} $ #1} 
\newcommand{\utwo}[1][]{ $ 0.03^{ + 0.36}_{ - 0.21} $ #1}

\newcommand{\tequi}[1][]{ $ 1102^{ + 26}_{ - 20} $ #1 }  
\newcommand{\pmass}[1][$M_{\oplus}$]{ $ 32.2 \pm 8.1 $ #1 }
\newcommand{\pden}[1][g\,cm$^{-3}$]{$ 2.15^{ + 0.67}_{ - 0.60}$ #1}

\newcommand\sme{\texttt{SME\,4.43}}
\newcommand\halpha{H$_\alpha$}                   
\newcommand\hbeta{H$_\beta$}
\newcommand\mgi{Mg\,{\sc i} }

\newcommand{\cai}{Ca\,{\sc i} }

\newcommand{\ngen}{{NextGen}} 		
\newcommand{\attw}{{ATLAS12}}           
\newcommand{\atni}{{ATLAS9}} 			
\newcommand{\marct}{{MARCS}} 		

\begin{document}

\shortauthors{Barrag\'an et al.}
\shorttitle{The Neptune-sized planet \pname}

\title{\pname: A 32-M$_\oplus$ Neptune-sized planet in a 10-day orbit transiting an F8 star}

\author{Oscar Barrag\'an\altaffilmark{1},
Sascha Grziwa\altaffilmark{2}, 
Davide Gandolfi\altaffilmark{1,3}, 
Malcolm Fridlund\altaffilmark{4,5},
Michael~Endl\altaffilmark{6},
Hans J. Deeg\altaffilmark{7,8}, 
Manuel~P.~Cagigal\altaffilmark{9},
Antonino F. Lanza\altaffilmark{10},
Pier G. Prada Moroni\altaffilmark{11,12},
Alexis M. S. Smith\altaffilmark{13},
Judith Korth\altaffilmark{2}, 
Megan~Bedell\altaffilmark{14},
Juan~Cabrera\altaffilmark{13}, 
William~D.~Cochran\altaffilmark{6},
Felice~Cusano\altaffilmark{15},
Szilard Csizmadia\altaffilmark{13}, 
Philipp~Eigm\"uller\altaffilmark{13}, 
Anders Erikson\altaffilmark{13}, 
Eike~W.~Guenther\altaffilmark{16}, 
Artie~P.~Hatzes\altaffilmark{16}, 
David~Nespral\altaffilmark{7,8}, 
Martin~P\"atzold\altaffilmark{2}, 
Jorge~Prieto-Arranz\altaffilmark{7,8},
and Heike~Rauer\altaffilmark{13,17}  
}

\altaffiltext{1}{Dipartimento di Fisica, Universit\`a di Torino, via P. Giuria 1, 10125 Torino, Italy; oscar.barraganvil@edu.unito.it}
\altaffiltext{2}{Rheinisches Institut f\"ur Umweltforschung an der Universit\"at zu K\"oln, Aachener Strasse 209, 50931 K\"oln, Germany}
\altaffiltext{3}{Landessternwarte K\"onigstuhl, Zentrum f\"ur Astronomie der Universit\"at Heidelberg, K\"onigstuhl 12, 69117 Heidelberg, Germany}
\altaffiltext{4}{Leiden Observatory, University of Leiden, PO Box 9513, 2300 RA, Leiden, The Netherlands}
\altaffiltext{5}{Department of Earth and Space Sciences, Chalmers University of Technology, Onsala Space Observatory, 439 92 Onsala, Sweden}
\altaffiltext{6}{Department of Astronomy and McDonald Observatory, University of Texas at Austin, 2515 Speedway, Stop C1400, Austin, TX 78712, USA}
\altaffiltext{7}{Instituto de Astrof\'isica de Canarias, 38205 La Laguna, Tenerife, Spain}
\altaffiltext{8}{Departamento de Astrof\'isica, Universidad de La Laguna, 38206 La Laguna, Spain}
\altaffiltext{9}{Departamento de F\'isica Aplicada, Universidad de Cantabria Avenida de los Castros s/n, E-39005 Santander, Spain}
\altaffiltext{10}{INAF - Osservatorio Astrofisico di Catania, via S. Sofia 78, 95123, Catania, Italy}
\altaffiltext{11}{Physics Department "E. Fermi", University of Pisa, Largo B. Pontecorvo 3, 56127, Pisa, Italy} 
\altaffiltext{12}{Istituto Nazionale di Fisica Nucleare, Largo B. Pontecorvo 3, 56127, Pisa, Italy}
\altaffiltext{13}{Institute of Planetary Research, German Aerospace Center, Rutherfordstrasse 2, 12489 Berlin, Germany}
\altaffiltext{14}{Department of Astronomy and Astrophysics, University of Chicago, 5640 S. Ellis Ave, Chicago, IL 60637, USA}
\altaffiltext{15}{INAF – Osservatorio Astronomico di Bologna, Via Ranzani, 1, 40127, Bologna}
\altaffiltext{16}{Th\"uringer Landessternwarte Tautenburg, Sternwarte 5, D-07778 Tautenberg, Germany}
\altaffiltext{17}{Center for Astronomy and Astrophysics, TU Berlin, Hardenbergstr. 36, 10623 Berlin, Germany}

\begin{abstract}

We report the discovery of \pname\ (\sname\,b), a transiting Neptune-sized planet monitored by the K2 mission during its campaign 5. We combine the K2 time-series data with ground-based photometric and spectroscopic follow-up observations to confirm the planetary nature of the object and derive its mass, radius, and orbital parameters. \pname\ is a warm Neptune-like planet in a 10-day orbit around a V=12.2~mag F-type star with $M_\star$=\smass\,, $R_\star$=\sradius, and age of $5.2_{-1.0}^{+1.2}$~Gyr. We derive a planetary mass and radius of $M_\mathrm{p}$=\pmass\ and $R_\mathrm{p}$=\pradius. \pname\ joins the relatively small group of Neptune-sized planets whose both mass and radius have been derived with a precision better than 25\,\%. We estimate that the planet will be engulfed by its host star in $\sim$3~Gyr, due to the evolution of the latter towards the red giant branch.

\end{abstract}


\section{Introduction} 
\label{sec:intro}

The transit of an exoplanet in front of its host star provides us with valuable information about its size. When combined with radial velocity (RV) measurements \citep[e.g.,][]{Mayor1995} or transit timing variations \citep[TTVs, e.g.,][]{2011ApJS..197....2F}, transit photometry gives us access to the geometry of the orbit, enabling the measurement of the true mass of the planet, of its radius, and consequently of its mean density. Masses, radii, densities, and orbital parameters are fundamental ``ingredients'' to study the internal structure, composition, dynamical evolution, tidal interaction, architecture, and atmosphere of exoplanets \citep[e.g.,][]{2015ARA&A..53..409W, 2016SSRv..tmp...25H}.

The space-based photometry revolution of \corot\ \citep{Baglin2006} and \kepler\ \citep{Borucki2010} has given us access to the small-radius planet domain ($R_\mathrm{p}$\,$\lesssim$\,$6\,R_{\oplus}$, i.e., Neptune- and Earth-sized planets), a regime that is not easily accessible from the ground. Neptune-like planets \citep[2.0\,$\lesssim$\,R$_{\mathrm{p}}$\,$\lesssim$\,6.0\,R$_{\oplus}$, 10\,$\lesssim$\,M$_{\mathrm{p}}$\,$\lesssim$\,40\,M$_{\oplus}$][]{2011ApJ...728..117B} are of special interest as they mark the transition from Super-Earths to larger planets with higher volatile content, more akin to the icy giants in our solar system. However, our knowledge of these planets is still quite limited. Although {\it Kepler} has found that $\sim$26\,\% of Sun-like stars in our Galaxy host small planets with orbital period shorter than 100 days \citep[][]{2014PNAS..11112655M}, determinations of masses with a precision of $\sim$25\,\% -- or better -- have been possible only for a few dozen Neptune-like planets\footnote{As of June 2016; source: \url{exoplanet.org}.}. This is because of the small RV variations induced by such planets and the faintness of most of the {\it Kepler} host stars (V$>$13~mag), which makes them not suitable for precise RV follow-up observations. 

In its extended K2 mission, \kepler\ is surveying different stellar fields located along the ecliptic,  performing 80-day-long continuous observations of 10\,000--20\,000 stars per campaign. K2 data products have no proprietary period and are released to the community typically three months after the end of each campaign, enabling immediate follow-up observations. The K2 mission is an unique opportunity to gain knowledge of transiting Neptune-sized planets  \citep[e.g.,][]{2016arXiv160107608E,2016Natur.534..658D}.
K2 is targeting a number of bright dwarfs (V$\leq$12\,mag) higher than the original \kepler\ mission \citep{Howell14}. This is a definitive advantage for any RV follow-up observations.

As part of the KEST, ESPRINT, and PICK2 collaborations \citep{2015DPS....4741702C,Grziwa15,2015ApJ...812..112S,2016AJ....151..171J, 2016arXiv160401265N}, we have recently started a RV follow-up program that aims at confirming Neptune-sized candidates detected by the K2 mission and at measuring their masses via high-precision RV follow-up observations. We herein report the discovery of \object{\pname} (\object{EPIC\,211391664\,b}), a transiting Neptune-sized planet in a 10-day orbit around a relatively bright (V=12.2 mag) solar-like star photometrically monitored by the K2 mission during its Campaign~5. We combine the K2 photometry with ground-based follow-up observations to assess the planetary nature of the transiting object and derive its mass. 
We note that \pname\ has been recently identified as a planet candidate by \cite{2016MNRAS.tmp.1017P} and \cite{2016arXiv160702339B}, but has not previously been confirmed. We are the first team to confirm and characterize in detail this planetary system.

The paper is organized as follows: in Sect.~\ref{sec:lcurve} we present the K2 photometry, and in Sect.~\ref{sec:luckyi} and \ref{sec:obts} our ground-based photometric and spectroscopic follow-up, respectively. Sect.~\ref{sec:spect} reports on the characterization of the host star. Sect.~\ref{sec:rvtransit} describes the joint RV and photometric analysis. Results, discussion, and conclusion are given in Sect.~\ref{sec:results} and~\ref{sec:conc}. 

\section{K2 light curve}  
\label{sec:lcurve}

K2 Campaign 5 observations began on 27 April 2015 UT and lasted until 10 July 2015 UT\footnote{See \url{http://keplerscience.arc.nasa.gov/k2-fields.html}.}. 
During the observations the boresight of the {\it Kepler} spacecraft was pointed at coordinates $\alpha$\,=\,$08^{\mathrm{h}}\,40^{\mathrm{m}}\,38^{\mathrm{s}}$, $\delta$\,=\,$+16^{\circ}\,49\,^\prime47\,^{\prime\prime}$. A total of 26\,054 light curves were simultaneously acquired by K2; 25\,850 in long cadence mode ($\sim$30 minute integration time) and 204 in short cadence mode ($\sim$1~minute integration time).

In this work, we use the light curves extracted by \cite{VanderburgJohnson14}\footnote{Publicly available at \url{https://www.cfa.harvard.edu/~avanderb/allk2c5obs.html}.}. They were the only publicly available light curves at the time we started the detection of transiting planet candidates in K2 Field 5.  We search the light curves for transit signals using the \texttt{DST} algorithm \citep{Cabrera12} and the \texttt{EXOTRANS} pipeline \citep{Grziwa12}. \texttt{DST} and \texttt{EXOTRANS} have been applied extensively to both \corot\ \citep{Carpano09,Cabrera09,Erikson12,Carone12,Cavarroc12} and \kepler\ \citep[][]{Cabrera14,2016arXiv160708417G} data. 
All transit detection algorithms search for a pattern in the data and use statistics to assess whether a signal is present in the data or not. When compared to widely used algorithms such as, e.g., Box Least Squares \citep[\texttt{BLS};][]{Kovacs02}, \texttt{DST} uses an optimized transit shape, with the same number of free parameters as \texttt{BLS}, and an optimized statistic for signal detection. \texttt{EXOTRANS} uses a combination of the wavelet based filter technique \texttt{VARLET} \citep{2016arXiv160708417G} and the \texttt{BLS} detection algorithm. \texttt{VARLET} was developed to reduce both stellar variability and data discontinuities. EXOTRANS calculates the Signal Detection Efficiency (SDE) for every light curve when the BLS algorithm is used. The Generalized Extreme Value (GEV) distribution is used to calculate the SDE threshold \citep{Grziwa12}. We consider all light curves with a SDE value higher than the SDE threshold for further inspection (about 4\,\% of the sample).

Both \texttt{DST} and \texttt{EXOTRANS} identify a periodic transit-like signal associated with the target \sname. The star was proposed for K2 observations by programs GO5007 (P.I.: J. Winn) and GO5029 (P.I.: D. Charbonneau). 
For brevity we will hereafter refer to the star and its transiting planet as \ssname\ and \pname, respectively. 

The target passes all of the tests that we carry out to identify likely false positives with the \texttt{DST} and \texttt{EXOTRANS} pipelines. These tests were regularly used during the \corot\ mission. Briefly, we stack and fit even and odd transits separately using the Transit Analysis package \texttt{TAP} \citep{Gazak2012}. We find neither significant odd-even transit depth variations, nor ellipsoidal variability/tidal deformation signatures in the light curve, both typically observed in eclipsing binaries. We find also no shallow secondary eclipses that might suggest an eclipsing binary scenario. Possible secondary eclipses are simulated using the detached eclipsing binary light curve fitter \citep[\texttt{DEBIL};][]{Devor2005} we first described in \cite{Patzold2012}. Similar tests are performed using the \texttt{DST} pipeline and are described in \cite{Cabrera09} and \cite{Cabrera12}. Large photometric variation in phase with the candidate orbital period is a hint for a possible binary. Such variations are also not found, and so we proceed to more detailed fitting of the light curve, as well as high-resolution imaging, reconnaissance spectroscopy, and RV observations (Sect.~\ref{sec:luckyi} and \ref{sec:obts}). 

We also search the K2 light curve of \ssname\ for additional transit signals, but none are found. The main identifiers, optical and infrared magnitudes, and proper motions of this star are listed in Table~\ref{tab:parstellar}.  

\begin{table}
\label{tab:parstellar}
\tabletypesize{\scriptsize}
\tablecolumns{3}
\tablewidth{0pt}
\caption{Main identifiers, magnitudes, and proper motion of \ssname.}
\begin{center}
\begin{tabular}{lcc}
\hline
\hline
\noalign{\smallskip}
Parameter & Value &  Source \\
\noalign{\smallskip}
\hline
\noalign{\smallskip}
\multicolumn{3}{l}{\emph{Main Identifiers}} \\
\noalign{\smallskip}
EPIC & 211391664  & EPIC \\
UCAC & 508-047859 & EPIC \\
2MASS & 08255719+1130402 & EPIC \\
$\alpha$(J2000.0) & $08^\mathrm{h}05^{\mathrm{m}}57.189^{\mathrm{s}}$ & EPIC \\
$\delta$(J2000.0) & +11$^{\circ}$30$^\prime$40.12${\arcsec}$ & EPIC \\
\noalign{\smallskip}
\hline
\noalign{\smallskip}
\multicolumn{3}{l}{\emph{Magnitudes}} \\
$B$ & 12.646$\pm$0.030 & EPIC \\
$V$ & 12.166$\pm$0.030 & EPIC \\
$g$ & 12.313$\pm$0.030 & EPIC \\
$r$ & 12.031$\pm$0.030 & EPIC \\
$J$ & 11.124$\pm$0.022 & 2MASS \\
$H$ & 10.905$\pm$0.025 & 2MASS \\
$K$ & 10.869$\pm$0.028 & 2MASS \\
$W1$ & 10.823$\pm$0.023 & WISE \\
$W2$ & 10.856$\pm$0.020 & WISE \\
$W3$ & 10.678$\pm$0.108 & WISE \\
$W4$ &  8.258           & WISE \\
\noalign{\smallskip}
\hline
\noalign{\smallskip}
\multicolumn{3}{l}{\emph{Proper motions}} \\
$\mu_{\alpha} \cos \delta$ (mas \ yr$^{-1}$) & $ -15.4 \pm 2.3 $ & UCAC2\\
$\mu_{\delta} $ (mas \ yr$^{-1}$) & $-8.8 \pm 1.5$ & UCAC2\\
\noalign{\smallskip}
\hline
\end{tabular}
\end{center}
\vspace{-0.5cm}
\tablecomments{Values of fields marked with EPIC are taken from the Ecliptic Plane Input Catalog, available at \url{http://archive. stsci.edu/k2/epic/search.php}. Values marked with UCAC2, 2MASS, and WISE are from \cite{2004AJ....127.3043Z}, \cite{Cutri2003}, \cite{Cutri2013}, respectively. The WISE $W4$ magnitude is an upper limit.}
\end{table}

\section{FASTACAM high-resolution imaging}
\label{sec:luckyi}

\begin{figure}
\plotone{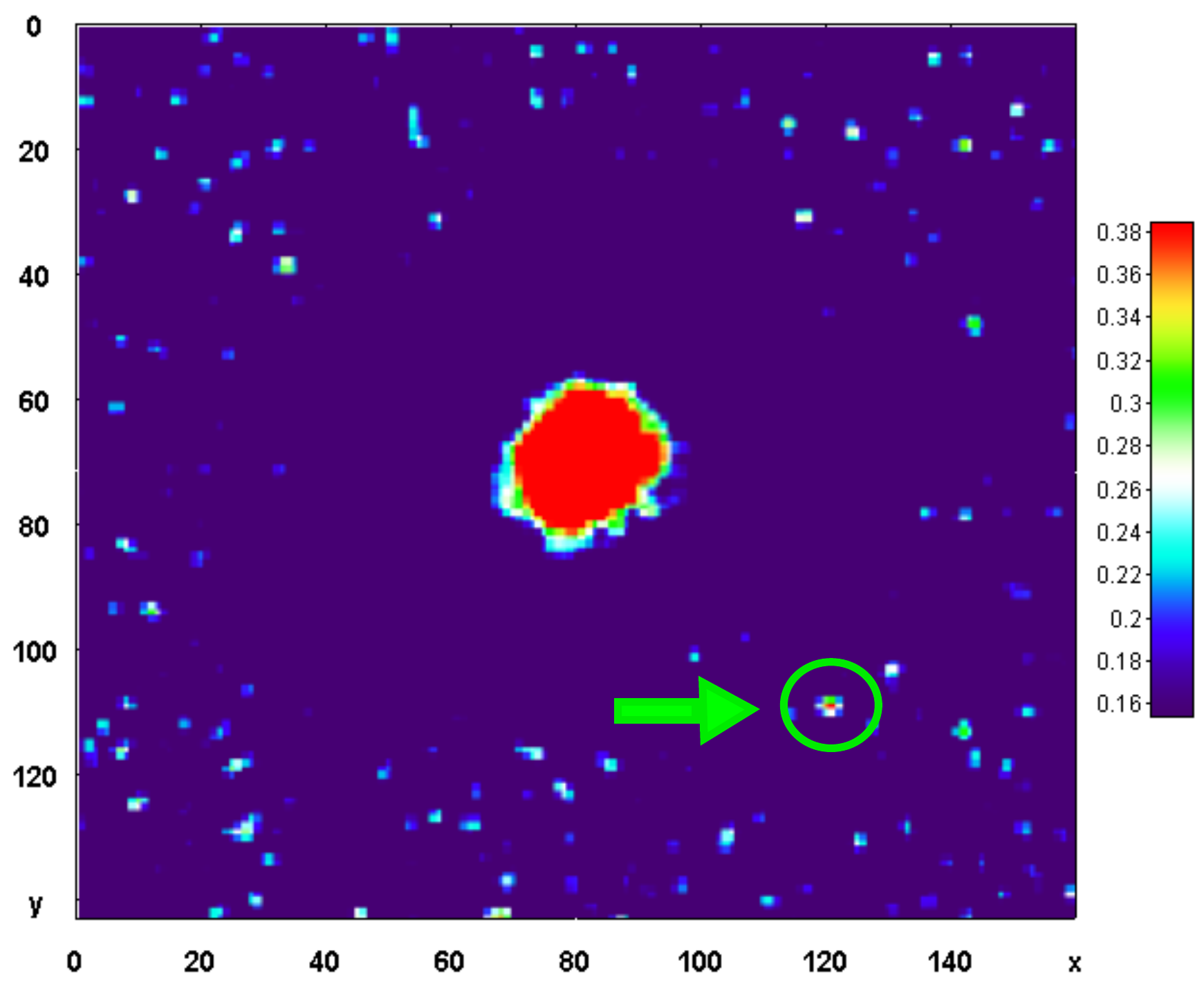}
\caption{FASTCAM image of \ssname\ processed with \texttt{COELI}. The pixel scale is 0.042\arcsec; North is left, East is down. \ssname\ is at the center of the ring-shaped feature, which is an artifact of the image processing. The faint nearby star 1.9\arcsec\ South-East of \ssname\ is indicated with a green arrow. Colors from blue to red represent the increasing level of temporal covariance with the central target, following the data processing with \texttt{COELI}.\label{fig:luckyi}}
\end{figure}

We observed \ssname\ on 17 May 2016 with the FASTCAM lucky imaging camera \citep{2008SPIE.7014E..47O} mounted on the 1.5m Carlos S\'anchez Telescope of Teide Observatory in Tenerife (Spain). To account for the low altitude of the object at the time of our observation, we used a relatively long exposure time of 300 milliseconds and acquired a total of 5\,000 images. The integration time of 300 milliseconds does not completely freeze the atmosphere, but this duration was necessary to collect enough light to detect faint objects. We selected the 300 best images, i.e., those with the highest Strehl ratio, and processed the data using the \texttt{COELI}\footnote{ImageJ Plugin available at \url{https://imagej.nih.gov/ij/plugins/index.html}.} algorithm \citep{2016MNRAS.455.2765C}. \texttt{COELI} provides a map of the temporal covariance between the intensity of \ssname\ and the intensity of the remaining pixels (Fig.~\ref{fig:luckyi}). This removes the speckled halo surrounding the host star and creates a dark ring-shaped region around it, which is the zone were the algorithm is more sensitive to the presence of faint objects \citep{2016MNRAS.tmp.1474C}. \texttt{COELI} also reinforces in this zone those pixels whose intensity follows the same temporal fluctuations as \ssname, which can only happen when the pixels contain an object. We estimate that in the ring-shaped region, at distances of 0.5--1.7\arcsec, there are no background objects brighter than V$\approx$19 mag (i.e., $\Delta$V$\approx$7 mag).

The final image shows the target to be isolated except for the detection of an object located 1.9\arcsec\ South-East of \ssname. The detected object is located just outside the dark ring-shaped region, in a zone where \texttt{COELI} provides relatively poor contrast and small spots show arbitrarily amplified noise. Nevertheless, due to being rather bright, we consider the source as a secure detection, and estimate it to be 50$\pm$10 times fainter (4.2$\pm$0.2~mag) than the main target. The distance of 1.9\arcsec\ between the target and the faint object is less than the sky-projected size of the {\it Kepler}/K2 CCD pixel ($\sim$4\arcsec). We therefore assume that the light from the faint object contributes with a fraction of 1/(50$\pm$10) to the measured flux of \ssname\ and correct the K2 light curve accordingly prior to performing the joint analysis presented in Sect.\ref{sec:rvtransit}.

No additional contaminants are identified. The DSS images reveal that the next closest star with a brightness comparable to the target is a $\sim$3.5-mag fainter object that is located at $\sim$42\arcsec\ East of \ssname, which is too large to produce any relevant influence onto the K2 light curve of \ssname.

\section{Spectroscopic Follow-up Observations} 
\label{sec:obts}

We took 2 reconnaissance spectra of \ssname \ with the Harlan J. Smith 2.7m Telescope and the Tull Coud\'e Spectrograph \citep{1995PASP..107..251T} at McDonald Observatory. The Tull spectrograph covers the entire optical spectrum (3450--9800\,\AA) at a resolving power of R$\approx$60\,000. We used exposure times of 1800 seconds, which resulted in a signal-to-noise ratio (SNR) of $\sim30$ per pixel at 5500~\AA. We derived a first estimate of the spectroscopic parameters by using our code \texttt{Kea} that compares observed high-resolution spectra to a large library of synthetic models \citep{2016arXiv160408170E}. For the first spectrum we obtain the following parameters: \teff\,$=5880\pm107$~K,  \logg\,$=3.81\pm0.31$ (cgs), [Fe/H]\,$=-0.06\pm0.07$~dex and a \vsini\,$=8.8\pm0.3$\,\kms. For the second observation: \teff\,$=5820\pm116$~K, [Fe/H]\,$=-0.03\pm0.08$~dex, \logg\,$=4.00\pm0.35$ (cgs) and a \vsini\,$=8.7\pm0.4$\,\kms. We also measure an absolute RV of~$76.7\pm0.2$\,\kms\ by cross-correlating the data with spectra of the RV standard star \object{HD\,50692} \citep{Udry1999}.

We also acquired 4 high-resolution spectra (R$\approx$67\,000) in November 2015 and January 2016 using the FIbre-fed \'Echelle Spectrograph \citep[FIES;][]{Frandsen1999,Telting2014} mounted at the 2.56m Nordic Optical Telescope (NOT) at Roque de los Muchachos Observatory (La Palma, Spain). We adopted the observing strategy described in \citet{Buchhave2010} and \citet{Gandolfi2013,Gandolfi2015}, i.e., we took 3 consecutive exposures of 1200 seconds per observation epoch -- to remove cosmic ray hits -- and acquired long-exposed (T$_\mathrm{exp}$$\approx$35 seconds) ThAr spectra immediately before and after the three sub-exposures -- to trace the RV drift of the instrument. We reduced the data using standard IRAF and IDL routines. The signal-to-noise ratio (SNR) of the extracted spectra is $\sim$30 per pixel at 5500\,\AA. Radial velocity measurements were derived via multi-order cross-correlation with the RV standard star \object{HD\,50692} -- observed with the same instrument set-up as \ssname. They are listed in Table~\ref{rvs} along with the full-width at half maximum (FWHM) and bisector span (BIS) of the cross-correlation function (CCF).

\begin{table}
\caption{Radial velocity measurements of \ssname. \label{rvs}}
\begin{tabular}{lccccr}
\hline
\hline
BJD$_\mathrm{TDB}$ & RV & $\sigma_{\mathrm{RV}}$ & CCF FWHM  & CCF BIS \\
$-{2\,450\,000}$ & (km s$^{-1}$) & (km s$^{-1}$)  & (km s$^{-1}$) & (km s$^{-1}$) \\
\hline
\noalign{\smallskip}
\multicolumn{2}{l}{FIES} \\
7342.706590 & 76.6027 & 0.0082 & 15.1659 & 0.0390 \\
7344.744082 & 76.6243 & 0.0081 & 15.1694 & 0.0401 \\
7347.706247 & 76.6143 & 0.0082 & 15.1770 & 0.0430 \\
7394.699773 & 76.6130 & 0.0082 & 15.1790 & 0.0366 \\
\hline
\noalign{\smallskip}
\multicolumn{2}{l}{HARPS} \\
7509.495449 & 76.7558 & 0.0084 & 10.3567 & 0.0430 \\
7511.495428 & 76.7516 & 0.0053 & 10.3617 & 0.0442 \\
7512.472984 & 76.7376 & 0.0055 & 10.3632 & 0.0450 \\
7516.525371 & 76.7425 & 0.0068 & 10.3729 & 0.0600 \\ 
\hline
\noalign{\smallskip}
\multicolumn{2}{l}{HARPS-N} \\
7371.582060 & 76.7393 & 0.0077 & 10.3474 & 0.0407 \\
7371.601679 & 76.7324 & 0.0098 & 10.3309 & 0.0499 \\
7448.440479 & 76.7499 & 0.0074 & 10.3352 & 0.0456 \\
7512.429111 & 76.7313 & 0.0048 & 10.3558 & 0.0580 \\
\hline
\end{tabular}
\end{table}

We also acquired 8 high resolution spectra using the HARPS \citep[R$\approx$115\,000;][]{2003Msngr.114...20M} and HARPS-N spectrographs \citep[R$\approx$115\,000;][]{Cosentino2012} mounted at the ESO-3.6m telescope at La Silla observatory  (Chile) and the 3.58m Telescopio Nazionale Galileo (TNG) at the Roque de los Muchachos Observatory (La Palma, Spain), respectively. The observations were performed between December 2015 and May 2016, setting the exposure times to 1800--3600 seconds depending on the sky condition. We monitored the Moon background light using the second fiber and reduced the data with the dedicated HARPS and HARPS-N data reduction software pipeline. The SNR of the extracted spectra is SNR=35--45 per pixel at 5500~\AA. Radial velocities (Table~\ref{rvs}) were extracted by cross-correlation with a G2 numerical mask \citep{Baranne96,Pepe02}.

We search for possible correlations between the RVs and the CCF FWHM, as well as between the RVs and CCF BIS. By combining all the three data sets, we derive for the RV and BIS data a Pearson correlation coefficient of -0.36 with a p-value of 0.25, while for the RV and FWHM measurements we obtain a Pearson correlation coefficient of -0.32 with a p-value of 0.31. The lack of significant correlations at a 0.05 confidence level provides further evidence that the observed RV variations are caused by the orbital motion of the planet rather than stellar activity. It also excludes the presence of an unseen stellar contaminant whose CCF is blended with the cross-correlation function of \ssname. We also perform a visual inspection of the Tull, FIES, HARPS, and HARPS-N spectra and search the CCFs for the presence of a secondary peak. We find no significant evidence of a second set
of spectral lines in the data.

\section{Properties of the host star}
\label{sec:spect}

We co-add the spectra from the NOT, ESO-3.6m, and TNG separately to get a combined FIES spectrum, a separate combined HARPS spectrum, and a third separate HARPS-N spectrum. The co-added data have a SNR of $\sim$100 per pixel at 5500~\AA. We use the three combined spectra to refine the estimates of the spectroscopic parameters of the host star. Following the spectral analysis of \corot\ and \kepler\ host stars \citep[e.g.,][]{2010A&A...512A..14F, Gandolfi2010, Gandolfi2015}, we select spectral features that are sensitive to different photospheric parameters. Our method is based on Spectroscopy Made Easy (\texttt{SME}), a software package that calculates synthetic spectra and fits them to high resolution observed spectra \citep{1996A&AS..118..595V}. \texttt{SME} is especially designed to determine basic stellar and atomic parameters from a match of the observed and normalized spectrum to the synthetic spectra generated from the parameterized atmospheres. It uses a non-linear least squares algorithm to solve for any subset of allowed parameters, which include atomic data (log\,$gf$ and van der Waals damping constants), the model atmosphere parameters (\teff, \logg), the metal abundances, and the projected rotational velocity \vsini. The \sme\ distribution includes a grid with a very large set of 1D-LTE plane-parallel stellar atmospheric models \citep[\atni, \attw, \ngen, and \marct\ models;][]{1993KurCD..13.....K,2013ascl.soft03024K,1999ApJ...512..377H,2008A&A...486..951G}. \attw\ is an opacity sampling model atmosphere program that computes the same models as \atni\ but instead of using pretabulated opacities and models with arbitrary abundances, \attw\ uses individual abundances and line data.

Our spectral analysis begins by primarily using the wings of the \halpha\ and \hbeta\ Balmer lines to determine \teff, adopting the calibration equations of \cite{2010MNRAS.405.1907B} and \cite{2014MNRAS.444.3592D} to estimate the microturbulent (\vmic) and macroturbulent (\vmac) velocities. The projected rotational velocity \visi\ is determined from a set of iron lines after which \mgi lines at $\lambda$\,=\,5167, 5173, and 5184~\AA\ and \cai lines at $\lambda$\,=\,6102, 6122, 6162, and 6439~\AA, are used to estimate the surface gravity \logg. In order to verify the accuracy of this method, we analyze a Solar spectrum from \cite{2011ApJS..195....6W}. Comparing with the discussion given in \cite{2005ApJS..159..141V}, we find the errors quoted there to be representative of what can currently be achieved when calculating synthetic spectra in order to fit high resolution, high SNR spectra. 

We obtain stellar parameters from the FIES, HARPS, HARPS-N consistent to within 1-sigma uncertainties. Our final adopted values for \teff, \logg, \met, and \vsini\ are the weighted means of the values produced by the three co-added spectra and the quoted errors are the 1-$\sigma$ standard deviation. They are also consistent within 2-$\sigma$ with the preliminary values derived from the 2 reconnaissance spectra taken at McDonald observatory (Sect.~\ref{sec:obts}). We note that the \vsini\ estimates obtained from the Tull spectroscopic data using \texttt{KEA} should be regarded as upper limits as they do not account for the line broadening induced by the macroturbulent velocity \citep{2016arXiv160408170E}.

We determine stellar mass, radius, and age by combining the effective temperature \teff\ and metallicity [M/H] with the mean density $\rho_\star$ obtained from the transit light curve modeling (Sect.~\ref{sec:rvtransit}). We compare the position of the host star on a $\rho_\star$-versus-\teff\ with a fine grid of evolutionary tracks. The latter are computed \emph{ad hoc} for this work using the \texttt{FRANEC} code \citep{Tognelli2011}, setting the same configuration as for the Pisa stellar evolution data base for low-mass stars\footnote{Available at \url{http://astro.df.unipi.it/stellar-models/}.} \citep{DellOmodarme2012}. We adopt the mixing-length parameter $\alpha_{ml}= $1.74, which is our solar calibrated value for the heavy element mixture of the Sun by \citet{Asplund2009}. We account for microscopic diffusion by means of the routine developed by \citet{Thoul1994}. The final grid contains tracks in the mass range 0.90-1.30~$M_{\sun}$, with a step of 0.01~$M_{\sun}$, computed for five different couples of initial metallicity $Z$ and helium abundance $Y$, namely, (0.006, 0.260), (0.008, 0.265), (0.010, 0.268), (0.011, 0.271), (0.012, 0.273), and (0.013, 0.274). We find that evolutionary models with initial metal content between Z=0.011 and Z=0.013 reproduce the current photospheric metallicity.  With a mass of $M_\star$=$1.074\pm0.042$~\Msun, radius of $R_\star$=$1.311^{+0.083}_{-0.048}$~\Rsun\, and an age of $5.2^{+1.2}_{-1.0}$~Gyr (Table~\ref{parstable}), \ssname\ is a slightly evolved star leaving the main sequence. Based on the calibration of \citet{Straizys81} for dwarf stars, the effective temperature of the star translates into a F8\,V spectral type. The stellar mass and radius imply a surface gravity of \logg\,$=4.23^{+0.03}_{-0.05}$~(cgs), which agrees within 1-$\sigma$ with the value of \logg\,$=4.35\pm0.10$~(cgs) derived from the co-added spectra.

We check the K2 data for evidence of rotational modulation. The lack of significant periodic and quasi-periodic photometric variation prevents us from estimating the stellar rotation period. Assuming that the star is seen equator-on, the projected rotational velocity \vsini=6.1$\pm$0.5~\kms\ and stellar radius $R_\star$=$1.311^{+0.083}_{-0.048}$\,\Rsun\ imply a rotational period of  $P_\mathrm{rot}$=10.9$^{+1.0}_{-0.8}$~days.

Following the technique described in \citet{Gandolfi2008}, we use the magnitudes listed in Table~\ref{tab:parstellar} and our spectroscopic parameters to estimate the interstellar extinction and distance to the star. We find that the light of \ssname\ suffers a negligible reddening ($A_\mathrm{v}$=0.05$\pm$0.05~mag) and the star is located at a distance $d=435^{+40}_{-20}$~pc from the Sun.

\section{Joint RV-Transit modeling} 
\label{sec:rvtransit}

We perform the joint modeling of the photometric and spectroscopic data using the code \texttt{pyaneti}, a Python/Fortran software suite that finds the best fitting solution using Markov Chain Monte Carlo (MCMC) methods based on Bayesian inference (Barrag\'an et al., in preparation). The code implements ensemble sampling with affine invariance for a larger coverage of parameter space \citep{2010CAMCO...5..65G}.

The photometric data included in the joint analysis are subsets of the whole K2 light curve. We select $\sim$13 hours of data-points centered on each of the 7 transits\footnote{The transit duration is $\sim$5 hours.} observed by K2. We de-trend the individual transits using a second-order polynomial locally fitted to the $\sim$16 out-of-transit points per transit (8 points per side). The final data-set contains 180 photometric points. The modeled RV data-set contains the 12 measurements listed in Table~\ref{rvs}.

The radial velocity model is given by a Keplerian orbit and an offset term for each systemic velocity \citep[see, e.g.,][]{2014exha.book.....P}. We fit for the systemic velocity $\gamma_j$ (as measured by the $j\mathrm{^{th}}$ instrument), the RV semi-amplitude variation $K$, the transit epoch $T_0$, the period $P_\mathrm{orb}$, the eccentricity $e$, and the argument of periastron of the star's orbit $\omega$ measured from the ascending node to its periastron.

The transit model follows the quadratic limb-darkened law of \cite{2002ApJ...580L.171M}. We account for the K2 long integration ($T_\mathrm{exp}$=29.425 minutes) by supers-ampling the transit model with 10 sub-samples per long cadence data \citep{Kipping2010}. For the linear $u_1$ and quadratic $u_2$ limb darkening coefficients, we use the $q_1=(u_1+u_2)^2$ and $q_2=u_1\,[\,2\,(u_1+u_2)\,]^{-1}$ parameterization described in \citet{2013MNRAS.435.2152K}. The fitted transit parameters are $T_0$, $P_\mathrm{orb}$, $e$, $\omega$, $q_1$, $q_2$, scaled semi-major axis $a/R_{\star}$, planet-to-star radius ratio $R_p/R_{\star}$, and impact parameter $b$. 

We use the Gaussian likelihood

\begin{equation}
\mathcal{L} =
(2 \pi)^{-n/2}
\left[
\prod_{i=1}^n \sigma_i
\right]
\exp
\left[
- \sum_{i=1}^n \frac{(D_i - M_i)^2}{2 \sigma_i^2 }
\right],
\end{equation}

where $n$\,=\,$n_\mathrm{tr}$\,+\,$n_\mathrm{RV}$ is the number of transit and RV points, and $\sigma_i$ is the error associated to each data point $D_i$, and $M_i$ is the model associated to a given $D_i$.

\begin{figure*}[t]
\includegraphics[height=2.25in]{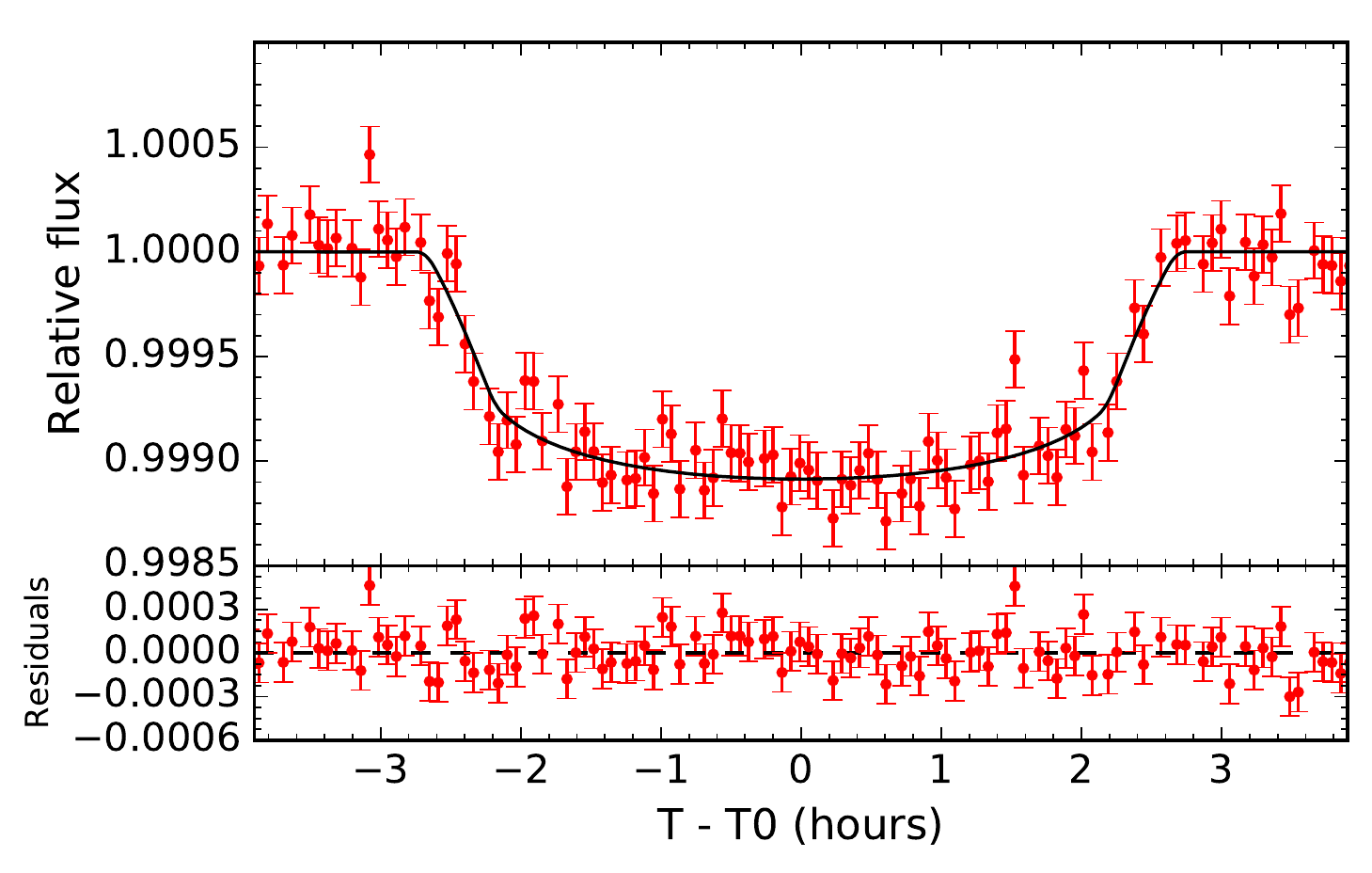}
\includegraphics[height=2.25in]{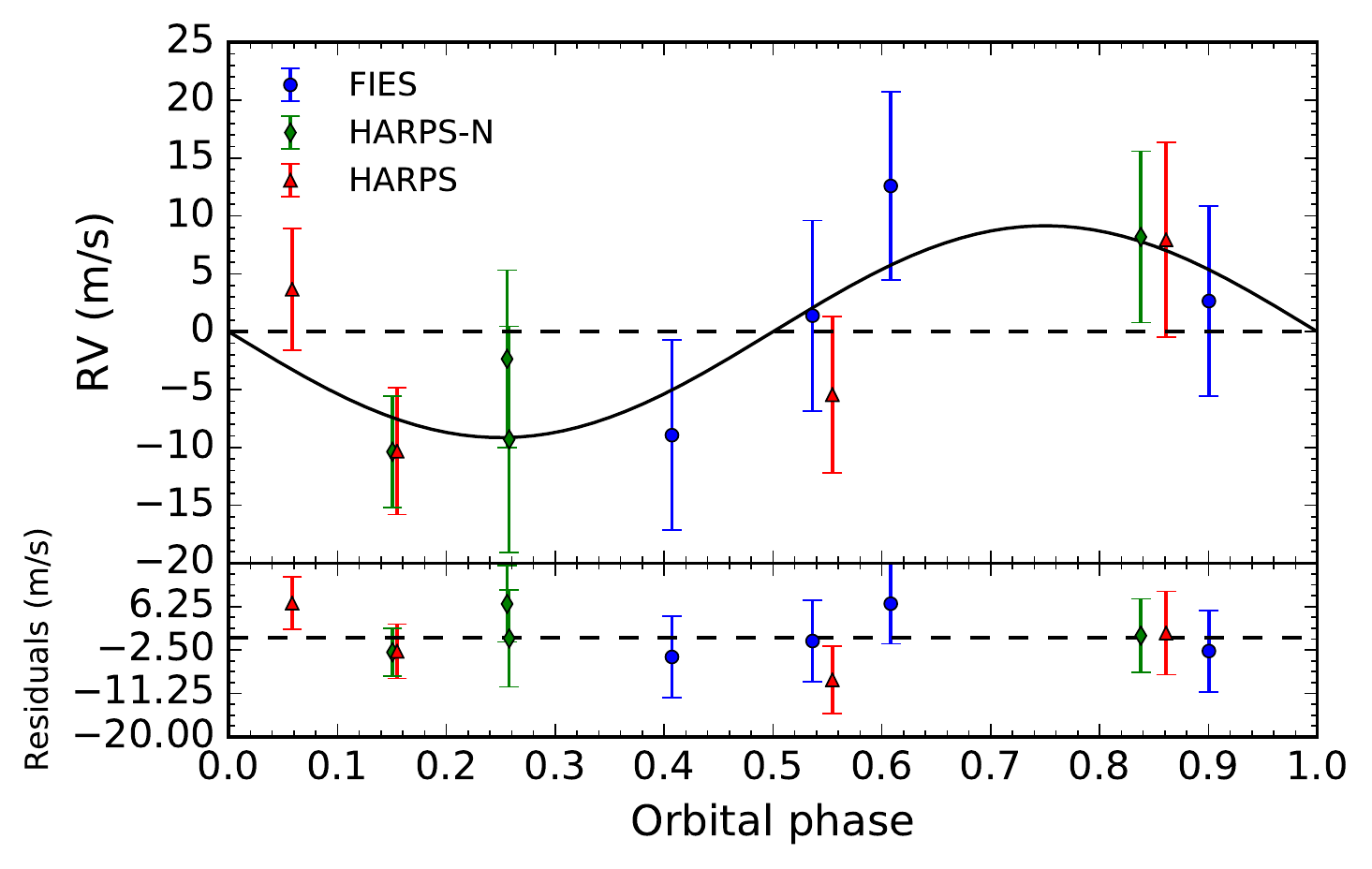}
\caption{\emph{Left panel}: Transit light curve folded to the orbital period of \pname\ and residuals. The red points are the K2 data and their error bars. The solid line mark the re-binned best fitting transit model. \emph{Right panel}: Phase-folded FIES (blue circles), HARPS-N (green diamonds) and HARPS (red triangles) RV measurements of \ssname\ and best fitting circular orbit (solid line), following the subtraction of the systemic velocities as measured from each instrument. \label{fig:fits}}
\end{figure*}

We fit for both a circular and an eccentric model. The joint modeling is carried out running 500 independent chains with uninformative uniform priors in the wide ranges $P_\mathrm{orb}$\,=\,$[10.1,\,10.2]$\,days, $T_0$\,=\,$[2457145.7,\,2457146.3]$, $b$\,=\,$[0,\,1]$, $a/R_{\star}$\,=\,$[5,\,100]$, $R_{\mathrm{p}}/R_{\star}$\,=\,$[0.005,\,0.2]$, $K$\,=\,$[0.001,\,1.0]$~\kms, and $\gamma_j$\,=\, $[1,\,100]$~\kms. For the circular model we set $e$\,=\,0 and $\omega$\,=\,$90 \deg$, while for the eccentric fit we set uniformative uniform priors between the limits $e$\,=\,$[0,\,1]$ and $\omega$\,=\,$[0,\,360] \deg$. For $q_1$ and $q_2$ we set uniformative uniform priors in the range $[0,\,1]$ to sample a physical solution for the limb darkening coefficients \citep{2013MNRAS.435.2152K}.

We check the chain convergence by comparing the ``between-chain'' and ``within-chain'' variance using the Gelman-Rubin statistics. The burning-in phase used 25\,000 additional iterations with a thin factor of 50, leading to a final number of 500 independent points for each chain, i.e., 250\,000 independent points for each fitted parameter.

An initial global fit to the data yields the parameterized limb darkening coefficients $q_1=$\qone\ and $q_2=$\qtwo\, which corresponds to $u_1=$\uone and $u_2=$\utwo. As described in \cite{2013A&A...549A...9C}, the large uncertainties arise from the shallow transit depth ($\sim$0.1\,\%), the small number of data points ($\sim$180) and transits (7), and the K2 long integration time ($\sim$30 minutes). We thus choose to constrain the limb darkening coefficient interpolating the table of \citet{Claret2011} and assuming conservative 20\,\% error bars. We stress that the system parameters derived with uninformative priors on the limb darkening coefficients are consistent to within 1-$\sigma$ uncertainties with those obtained by constraining $u_1$~and~$u_2$.

\section{Results and Discussion}\label{sec:results}

Figure \ref{fig:fits} shows the folded transit light curves and phase-folded RV curve, along with their best fitting models. The parameter estimates and error bars are listed in Table~\ref{parstable}. They are taken as the median and the 68\,\% central interval of the final posterior distributions \citep{2010blda.book.....G}. Our results are consistent with the transit parameters derived by \cite{2016MNRAS.tmp.1017P} and \cite{2016arXiv160702339B}.

Our RV measurements do not allow us to constrain the eccentricity of the system. A fit for an eccentric orbit yields $0.19^{+0.17}_{-0.13}$ with a significance of only about 1-$\sigma$. In order to further check whether the non-zero eccentricity solution is significant or not, we run an F-test and calculate the p-value, i.e., the probability that the apparent eccentricity could have arisen if the underlying orbit were circular \citep{Lucy1971}. In doing so we take into account the number of fitted parameters -- both for the circular and eccentric model~--, the number of measurements and their uncertainties, and the residuals from the best fitting circular and eccentric solution. We find a p-value of 0.87, which is much higher than the 0.05 significance threshold suggested by \citet{Lucy1971} to prefer $e$\,$\neq$\,0 over $e$\,=\,0. We therefore conclude that the nonzero best fitting eccentricity obtained with models where $e$ is allowed to vary is not significant. Moreover, we find that the circular (DOF=153) and eccentric (DOF=151) models provide very similar minimum $\chi^2$ values of $\sim$152. The difference of the Bayesian information criterion is $\Delta$BIC=10 between the two models, implying that the circular model is favored. We therefore adopt the circular model as the one that better describes our data. We note that the derived system parameters for a non-zero eccentricity are consistent to within 1-$\sigma$ uncertainties with those derived assuming a circular orbit.

\begin{figure}
\includegraphics[width=0.49\textwidth]{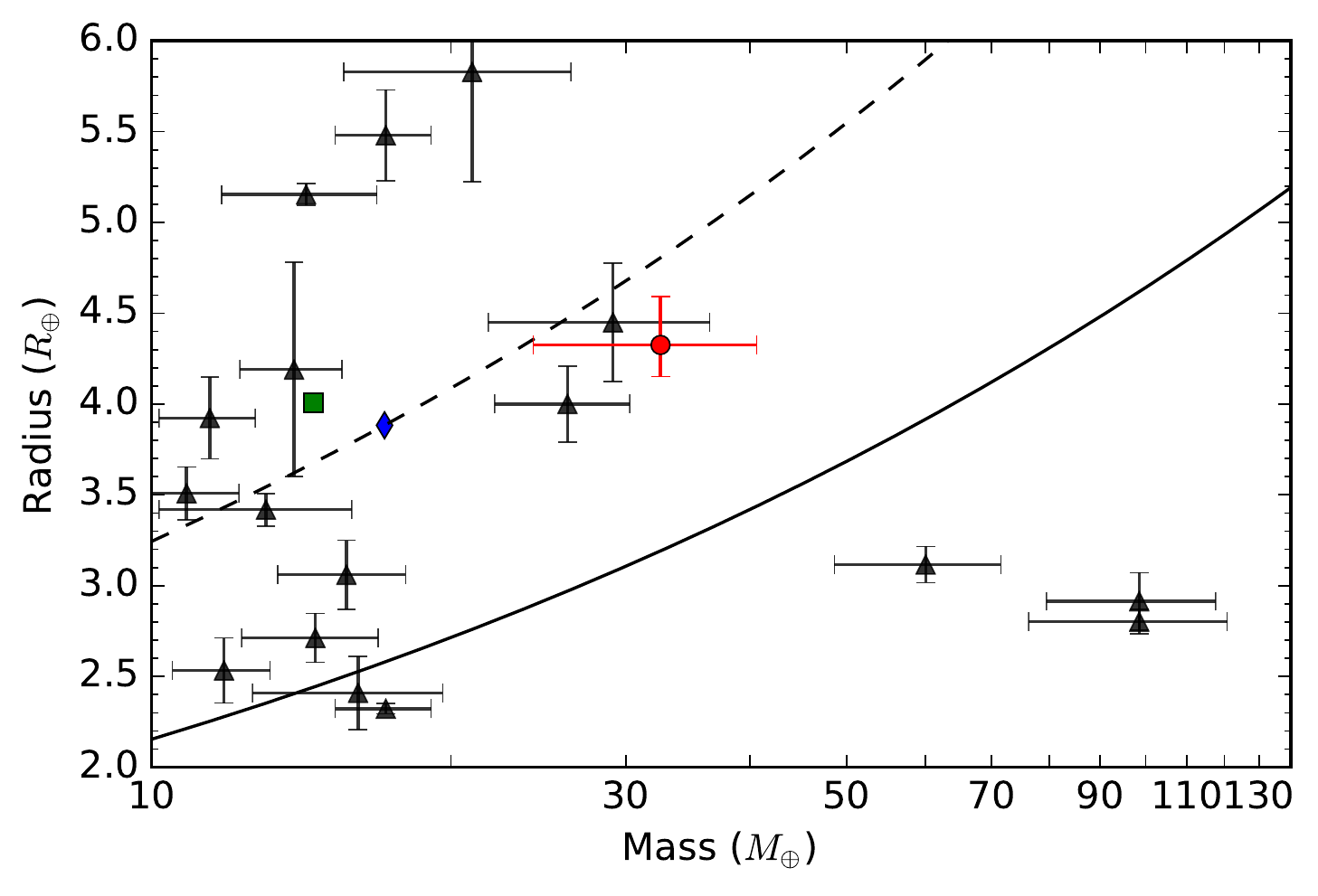}
\caption{Mass-radius diagram for Neptune-sized planets (2.0\,$\lesssim$\,$R_{\mathrm{p}}$\,$\lesssim$\,6.0\,$R_\oplus$) whose both mass and radius have been determined with a precision of at least $\sim$25\,\% \citep[Exoplanet Orbit Database, as of June 2016;][]{2014PASP..126..827H}. The red circle marks the position of \pname. The green diamond and blue square show the position of Neptune and Uranus, respectively. The solid and dashed lines mark the Earth (5.5~$\mathrm{g\,cm^{-3}}$) and Neptune (1.6~$\mathrm{g\,cm^{-3}}$) isodensity curves. \label{fig:f3} }
\end{figure}

\pname\ has a mass of $M_\mathrm{p}$\,=\,\pmass\ and a radius of $R_\mathrm{p}$\,=\,\pradius, consistent with a density of \pden. These parameters are calculated adopting the stellar mass and radius derived in Sect.~\ref{sec:spect} and listed in Table~\ref{parstable}. Figure \ref{fig:f3} shows the position of \pname\ in the mass-radius diagram for Neptune-sized planets. The plot includes only those objects whose both mass and radius have been estimated with a precision of at least $\sim$25\,\%. \pname\ joins the family of intermediate mass (20$<$$M_\mathrm{p}$$<$50\,$M_\oplus$) Neptune-sized planets. Whereas its radius is slightly larger than that of Neptune (3.9\,$R_\oplus$), the mass of \pname\ is almost twice as large as the mass of Neptune. This implies that a solid massive core surrounded by a large atmosphere is expected \citep[see, e.g.,][]{2014ApJ...783L...6W}.

Assuming a minimum mass solar nebula (MMSN), the isolation mass \citep{2014ApJ...795L..15S} of a planet at 0.093 AU is $\sim$0.004\,$M_\oplus$, which is significantly lower than the mass of \pname. In order to form \pname\ \emph{in situ}, a disk surface density $\sim$5500 times larger than the MMSN is required. This value would generate gravitational instabilities in the disk, because its Toomre parameter would be $Q \approx 0.03 \ll 1$ \citep{2014ApJ...795L..15S}. This scenario does not support the \emph{in situ} formation of \pname.

\cite{2014ApJ...793L...3V} proposed that Neptune-mass planets may form via migration of hot Jupiters that come so close to their host stars as to fill their Roche lobe and start conservative mass transfer to the star. This may reverse the direction of migration and increase the orbital period. However, it seems very difficult to reach a final orbital period of about 10 days, as in the case of \pname. Moreover, this formation scenario cannot easily account for the measured relatively low density of the planet (\pden). Therefore, we argue that \pname\ likely formed in the outer region of the protoplanetary disk and then migrated inwards to its current position \citep[see, e.g.][]{2012ARA&A..50..211K}.

We integrate the equations of tidal and rotational evolution as in \cite{2016arXiv160608623L} assuming a constant modified tidal quality factor $Q_\star^{\prime}$ for the star. Given that the stellar rotation period is close to the orbital period (Sect.~\ref{sec:spect}), tidal dissipation by inertial waves inside the star is considered leading to a remarkably stronger tidal interaction than in the case of the equilibrium tide \citep{2007ApJ...661.1180O}. 
Therefore, we explore the evolution for three fixed values of $Q_\star^{\prime}$, i.e., $10^{5}$, $10^{6}$, and $10^{7}$, from the stronger to the weaker coupling. Following \cite{2011A&A...529A..50L}, we include the loss of angular momentum produced by the stellar magnetized wind considering a saturation regime for an angular velocity greater than $8\,\Omega_{\odot}$, where $\Omega_{\odot}$ is the present solar angular velocity. We assume that the orbit of the planet is circular, although the tidal interaction is so weak that any initial eccentricity could survive up to the present stage of the system evolution (see below).

Fig.\,\ref{fig:lanza} shows the evolution of the rotation period of the star (upper panel), semi-major axis of the planet's orbit (middle panel), and stellar radius (lower panel) as obtained from the evolutionary models presented in Sect.~\ref{sec:spect}. Tidal interaction is so weak that there is virtually no evolution of the orbital separation since the planet arrived at the present semi-major axis (Fig.\,\ref{fig:lanza}, middle panel). The rotation of the star is braked solely by the stellar wind with a completely negligible tidal exchange between the orbital and the spin angular momenta, and no dependency on the stellar tidal quality factor $Q_\star^{\prime}$, owing to the small mass of the planet and large separation. (Fig.\,\ref{fig:lanza}, upper panel). Under our model assumptions, we estimate that the star reached the zero age main sequence (ZAMS) with a rotation period of about 1.5~days.

The tidal evolution of the planet will become important in the future -- after $\sim$3~Gyr from now -- due to the increase of the stellar radius and rotational period of the star, leading to a rapid decay of the planet's orbit (Fig.\,\ref{fig:lanza}, middle panel).

\begin{figure}[t]
\plotone{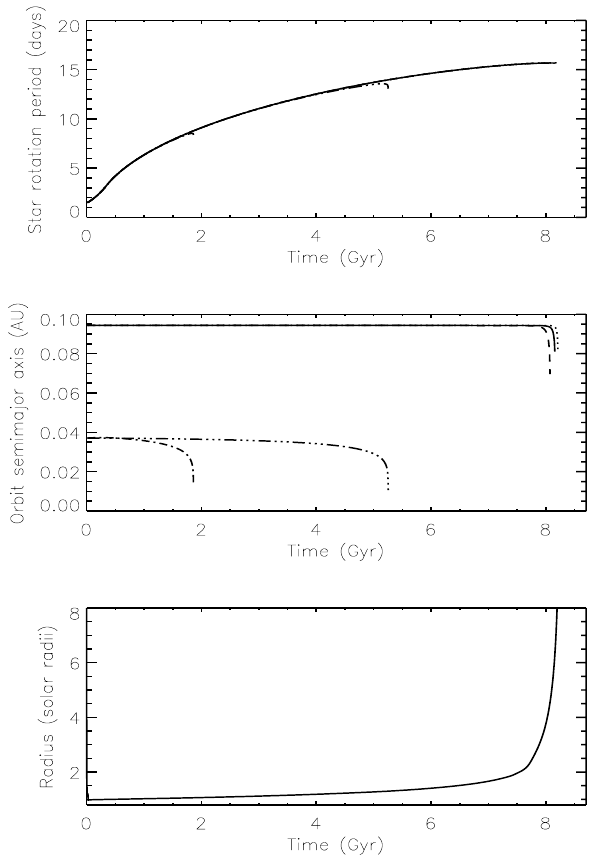}
\caption{Rotational period of the star (upper panel), semi-major axis  of the planet orbit (middle-panel), and stellar radius (lower panel) versus time. Different line styles refer to different initial semi-major axis $a_{0}$ and tidal quality factor of the star $Q_\star^{\prime}$ as follows: solid line: $Q_\star^{\prime} = 10^{6}$, $a_{0} = 0.0943$~AU; dotted line: $Q_\star^{\prime} = 10^{7}$, $a_{0} = 0.0943$~AU; dashed line: $Q_\star^{\prime} = 10^{5}$, $a_{0} = 0.0943$~AU; dash-dotted: $Q_\star^{\prime} = 10^{5}$, $a_{0} = 0.037$ AU~(corresponding to an orbital period of 2.5 days); dash-triple-dotted: $Q_\star^{\prime} = 10^{6}$, $a_{0} = 0.037$~AU. \label{fig:lanza}}
\end{figure}

The amount of angular momentum in the orbit is insufficient to synchronize the rotation of the star, so the present approximately synchronous state cannot be maintained. \cite{2015A&A...574A..39D} showed that other systems having host stars with an effective temperature around 6100~K show a rather wide distribution of the ratio of the orbital period to the stellar spin period, even in the case of more massive planets, thus supporting the conclusion that the present approximate synchronicity is probably coincidental.

Finally, we consider the possibility that the planet was initially significantly closer to the star when the latter reached the ZAMS and was pushed outwards by the action of tides because angular momentum was transferred from the stellar spin to the orbit, provided that the rotational period of the star was shorter than the orbital one. We find that also this scenario is unlikely. 
As an illustrative case, we show in Fig.~\ref{fig:lanza} two integrations for the planet initially at an orbital period of 2.5~days, corresponding to a semi-major axis of 0.037~AU. This is the minimum orbital period for observed Neptune-mass planets around main-sequence stars \citep[cf. Fig.~4 of ][]{2014ApJ...793L...3V} that we choose in order to maximize the strength of the tidal interaction. Since the star was initially rotating faster than the planet, the tidal interaction was initially pushing the planet outwards, in particular for $Q_\star^{\prime} = 10^{5}$ (Fig.~\ref{fig:lanza}, middle panel). However, the fast rotational braking of the star led soon to a rotation period longer than the orbital period. Since the amount of orbital angular momentum was too small to maintain the synchronous state, the final fate of the planet was to fall towards the star under the action of tides within a few~Gyrs\footnote{We note that assuming a different initial orbital period leads to qualitatively similar scenarios. If the initial orbital period of the planet is shorter than 2.5 days  (i.e., $a_{0} < 0.037$~AU), tidal push is stronger, but for a shorter time interval before the rotation period of the star becomes longer than the orbital period, after which  the orbit decays faster. If the planet is further out ($P_\mathrm{0,orb} > 2.5$~days and $a_{0} > 0.037$~AU), tides are weaker, but they can act longer before the direction of the evolution of the semi-major axis is reversed and the planet falls into the star.}. This scenario would account for the significant dearth of Neptune-like planets with orbital periods below 2-4 days \citep[see, e.g.,][]{2011ApJ...727L..44S,2016A&A...589A..75M}.

The tidal evolution of the system further supports an inward migration scenario for \pname, from the outer region of the system to its current position.

\vspace{-.25cm}
\section{Conclusions}\label{sec:conc}

We confirm the planetary nature of \pname\ and derive the system parameters. Our results are based on photometric data from the K2 space mission combined with high-precision Tull, FIES, HARPS, and HARPS-N RV measurements and lucky imaging. \pname\ is a transiting Neptune-sized planet in a 10-day-orbit around an F8\,V leaving the main sequence. It has a mass of $M_\mathrm{p}$\,=\,\pmass and a radius of $R_\mathrm{p}$\,=\,\pradius, translating into a mean density of \pden. \pname\ joins the still relatively small number of Neptune-size planets ($\sim$20 objects) whose mass and radius have been determined with a precision better than 25\,\%.
\newline

We thank the anonymous referee for the insightful, thoughtful and constructive review. We are very grateful to Simon~Albrecht for providing us with one of the HARPS-N measurement and for his valuable comments and suggestions. We are thankful to Jorge Melendez, Martin K\"urster, Fran\c{c}ois Bouchy, Nuno Santos, and Xavier Bonfils who kindly agreed to exchange HARPS time with us. We express our deepest gratitude to the NOT, ESO, TNG and McDonald staff members for their unique support during the observations. Szilard Csizmadia thanks the Hungarian OTKA Grant K113117. Hans Deeg and David Nespral acknowledge support by grant ESP2015-65712-C5-4-R of the Spanish Secretary of State for R\& D\&i (MINECO). This research was supported by the Ministerio de Economia y Competitividad under project FIS2012-31079. The research leading to these results has received funding from the European Union Seventh Framework Programme (FP7/2013-2016) under grant agreement No. 312430 (OPTICON) and from the NASA K2 Guest Observer Cycle 1 program under grant NNX15AV58G to The University of Texas at Austin. This paper includes data taken at McDonald Observatory of the University of Texas at Austin. Based on observations obtained \emph{a}) with the Nordic Optical Telescope (NOT), operated on the island of La Palma jointly by Denmark, Finland, Iceland, Norway, and Sweden, in the Spanish Observatorio del Roque de los Muchachos (ORM) of the Instituto de Astrof\'isica de Canarias (IAC); \emph{b}) with the Italian Telescopio Nazionale Galileo (TNG) also operated at the ORM (IAC) on the island of La Palma by the INAF - Fundaci\'on Galileo Galilei. Based on observations made with ESO Telescopes at the La Silla Observatory under programme ID 097.C-0948. This research made use of data acquired with the Carlos S\'anchez (TCS) Telescope, operated at Teide Observatory on the island of Tenerife by the Instituto de Astrof\'isica de Canarias. This paper includes data collected by the Kepler mission. Funding for the Kepler mission is provided by the NASA Science Mission directorate. This publication makes use of data products from the Two Micron All Sky Survey (2MASS), which is a joint project of the University of Massachusetts and the Infrared Processing and Analysis Center/California Institute of Technology, funded by the National Aeronautics and Space Administration and the National Science Foundation. This publication makes use of data products from the Wide-field Infrared Survey Explorer (WISE), which is a joint project of the University of California, Los Angeles, and the Jet Propulsion Laboratory/California Institute of Technology, funded by the National Aeronautics and Space Administration.

\facilities{Kepler (K2), NOT (FIES), ESO:3.6m (HARPS), Sanchez (FAST-CAM), Smith (Tull), TNG (HARPS-N).}

\software{\texttt{COELI}, \texttt{DEBIL}, \texttt{DST}, \texttt{EXOTRANS}, \texttt{Kea} \texttt{IDL}, \texttt{IRAF}, \texttt{pyaneti}, \texttt{SME}, \texttt{TAP}.}

\floattable
\begin{deluxetable}{lc}[t]
\tabletypesize{\scriptsize}
\tablecolumns{2}
\tablewidth{0pt}
\tablecaption{Stellar and Planetary Parameters.\label{parstable}}
\tablehead{
\colhead{Parameter} & \colhead{Value}}
\startdata
\multicolumn{2}{l}{\emph{Model Parameters}} \\
\noalign{\smallskip}
Orbital period $P_\mathrm{orb}$ (days) &  \porb[] \\
Transit epoch $T_0$ (BJD$_\mathrm{TDB}-$2\,450\,000) & \tzero[]  \\
Scaled semi-major axis $a/R_{\star}$ &  \saxis[]  \\
Scaled planet radius $R_\mathrm{p}/R_{\star}$  & \spradius[]  \\
Impact parameter, $b$    & \impactp   \\
Parameterized limb-darkening coefficient $q_1$\tablenotemark{a}  & \qonefixed \\
Parameterized limb-darkening coefficient $q_2$\tablenotemark{a}  & \qtwofixed \\
Eccentricity $e$ &  \ec[] (fixed) \\
Radial velocity semi-amplitude variation $K$ (m s$^{-1}$) & \krv[] \\
Systemic velocity $\gamma_{\mathrm{FIES}}$  (km s$^{-1}$) & \velF[] \\
Systemic velocity $\gamma_{\mathrm{HARPS}}$ (km s$^{-1}$)  & \velS[]  \\
Systemic velocity $\gamma_{\mathrm{HARPS-N}}$ (km s$^{-1}$) & \velH[] \\
\noalign{\smallskip}
\hline
\multicolumn{2}{l}{\emph{Derived parameters}} \\
Semi-major axis of the planetary orbit $a$ (AU) & \axis[]  \\
Transit duration $\tau_{14}$ (hours) & \ttotal[] \\
Transit ingress/egress duration $\tau_{12}=\tau_{34}$ (hours) &  \tineg[]  \\
Orbit inclination along the line-of-sight $i_\mathrm{p}$ ($^{\circ}$) & \inclination[] \\
\noalign{\smallskip}
\hline
\multicolumn{2}{l}{\emph{Stellar parameters}} \\
\noalign{\smallskip}
Star mass $M_{\star}$ ($M_\odot$) & \smass[] \\
Star radius $R_{\star}$ ($R_\odot$) &  \sradius[]  \\
Surface gravity \logg\ (cgs)\tablenotemark{b} &  $4.23^{+0.03}_{-0.05}$  \\
Mean density $\rho_{\star}$ (g cm$^{-3}$) & \sden[] \\
Star age (Gyr) & $5.2_{-1.0}^{+1.2}$  \\
Spectral type\tablenotemark{c}  & F8\,V  \\
Effective temperature $T_{\mathrm{eff}}$ (K) & $6120\pm80$  \\
Iron abundance $[\mathrm{Fe}/\mathrm{H}]$ (dex) &   $-0.2\pm0.1$  \\
Nickel abundance $[\mathrm{Ni}/\mathrm{H}]$  (dex) &   $-0.1\pm0.1$  \\
Silicon abundance $[\mathrm{Si}/\mathrm{H}]$  (dex) &   $-0.1\pm0.1$   \\
Calcium abundance $[\mathrm{Ca}/\mathrm{H}]$  (dex) &   $-0.1\pm0.1$   \\
Sodium abundance $[\mathrm{Na}/\mathrm{H}]$  (dex) &  $-0.0\pm0.1$ \\
Magnesium abundance $[\mathrm{Mg}/\mathrm{H}]$  (dex) &   $-0.0\pm0.1$ \\
Microturbulent velocity $v_{\mathrm{mic}}$\tablenotemark{c}&  $1.3\pm0.1$  \\
Macroturbulent velocity $v_{\mathrm{mac}}$\tablenotemark{d}  & $3.7\pm0.6$ \\
Projected rotational velocity $v\sin i_{\star}$  & $6.1\pm0.5$ \\
Distance $d$ (pc) & $435_{-20}^{+40}$   \\
Visual interstellar extinction $A_\mathrm{v}$ (mag) & $0.05\pm0.05$  \\ 
\noalign{\smallskip}
\hline
\multicolumn{2}{l}{\emph{Planetary parameters}} \\
\noalign{\smallskip}
Planet mass $M_\mathrm{p}$ ($M_\oplus$) & \pmass[]  \\
Planet radius $R_\mathrm{p}$ ($R_\oplus$) & \pradius[] \\
Planet density $\rho_\mathrm{p}$ (g\,cm$^{-3}$)  &  \pden[] \\
Equilibrium temperature $T_{\rm eq}$ (K)  &  \tequi[] \\
\noalign{\smallskip}
\enddata
\tablecomments{The adopted Sun and Earth units follow the recommendations from the International Astronomical Union \citep{2016arXiv160509788P}.} 
\tablenotetext{a}{The limb-darkening coefficient parameterization follows \citet{2013MNRAS.435.2152K}. The estimates have been obtained assuming $u_1=0.33\pm0.06$ and $u_2=0.30\pm0.06$ for the linear and quadratic limb-darkening coefficients \citep{Claret2011}, adopting 20\,\% conservative error bars.}
\tablenotetext{b}{Stellar surface gravity \logg\ as measured from the global fit and evolutionary tracks. The spectroscopic analysis gives \logg\,$=4.35\pm0.10$ (cgs).}
\tablenotetext{c}{Based on the spectral type vs. effective temperature calibration of \citet{Straizys81} for dwarf stars.}
\tablenotetext{d}{Micro and macroturbulent velocities from the calibration equations of \citet{2010MNRAS.405.1907B} and \citet{2014MNRAS.444.3592D}, respectively.}
\end{deluxetable}


\listofchanges

\end{document}